\documentclass[conference]{IEEEtran}
\IEEEoverridecommandlockouts
\usepackage{siunitx}
\usepackage{amsmath,amssymb,amsfonts}
\usepackage{algorithmic}
\usepackage{graphicx}
\usepackage{textcomp}
\usepackage{xcolor}
\usepackage{tabu}
\usepackage{array}
\usepackage{comment}
\newcolumntype{C}[1]{>{\centering\let\newline\\\arraybackslash\hspace{0pt}}m{#1}}
\usepackage{caption}
\usepackage{subcaption}
\usepackage{comment}
\usepackage{balance}
\usepackage{url}
\usepackage{hyperref}
\usepackage{cite}
\hypersetup{
    colorlinks=true,
    linkcolor=magenta,
    filecolor=blue,      
    urlcolor=cyan,
    citecolor=green,
}

\def\BibTeX{{\rm B\kern-.05em{\sc i\kern-.025em b}\kern-.08em
    T\kern-.1667em\lower.7ex\hbox{E}\kern-.125emX}}

\makeatletter
\def\ps@IEEEtitlepagestyle{%
    \def\@oddfoot{\mycopyrightnotice}%
    \def\@evenfoot{}%
}
\def\mycopyrightnotice{%
    {\footnotesize  978-0-7381-1102-5/20/\$31.00 ©2020 IEEE \hfill}
    \gdef\mycopyrightnotice{}
}


\makeatletter
\newcommand*\titleheader[1]{\gdef\@titleheader{#1}}
\AtBeginDocument{%
  \let\st@red@title\@title%
  \def\@title{%
    \bgroup\normalfont\large\raggedright\@titleheader\par\egroup
    \vskip.4em\st@red@title}
}
\makeatother    
\DeclareRobustCommand*{\IEEEauthorrefmark}[1]{%
  \raisebox{0pt}[0pt][0pt]{\textsuperscript{\footnotesize\ensuremath{#1}}}}
  
\title{ Android Based Low Cost Sitting Posture Monitoring System}
\titleheader{2020 11th International Conference on Electrical and Computer Engineering (ICECE)}
 \author{\IEEEauthorblockN{Nusrat Binta Nizam\IEEEauthorrefmark{1},Tohfatul Jinan\IEEEauthorrefmark{2}, Wahida Binte Naz Aurthy\IEEEauthorrefmark{3}, Md. Rakib Hossen\IEEEauthorrefmark{4} and Jahid Ferdous\IEEEauthorrefmark{5} \IEEEauthorblockA{Department of Biomedical Engineering, \\
 Bangladesh University of Engineering and Technology\\
Dhaka-1205, Bangladesh\\
 Email: \IEEEauthorrefmark{1}nusratbintanizam@ug.bme.buet.ac.bd,
 \IEEEauthorrefmark{2}tohfatuljinan@ug.bme.buet.ac.bd,
 \IEEEauthorrefmark{3}wahidabintenaz@ug.bme.buet.ac.bd,\\
 \IEEEauthorrefmark{4}mdhossen@ug.bme.buet.ac.bd,
 \IEEEauthorrefmark{5}ferdousj@bme.buet.ac.bd}}}

\begin{document}
\maketitle
\begin{abstract}
Back pain is one of the leading causes of disability-adjusted life year globally and the most common cause of low back pain is poor sitting posture. There are several actions that can be adopted proactively to avoid poor sitting posture induced back pain including behavioral change, regular exercise, and use of an ergonomic chair. However, these are either expensive and/or difficult to execute for prolonged periods. Sitting posture monitoring systems continuously observe the sitting pattern of a person in real-time and give feedback/alert poor sitting posture is observed.  In this study, a real-time posture monitoring system has been designed and a functional prototype has been developed using simple electrical elements and an android application. Appropriate position of the sensor in the spine to measure the degree of bending and the threshold sensor values for good posture sittings have been determined based on the results from healthy volunteers of different ages and height. The android application continuously monitors the degree of bending and provides vibration when the bending reaches the threshold of bad posture or the duration of the sitting crosses the clinically recommended time limit to prevent prolonged sitting. Positive user feedback has been received in terms of comfortability, effectiveness, and satisfaction levels. The manufacturing cost of the developed monitoring system is minimal compared to the available expensive systems in the market and the cost would further go down if it is produced in bulk. This device efficiently monitors the sitting posture pattern to prevent back pain and within the affordable price range for the people from middle to under-developed countries.
%%%%
\end{abstract}

\begin{IEEEkeywords}
Back pain; sitting posture monitoring system; android application; spine.

\end{IEEEkeywords}

\section{Introduction}
%Proper sitting posture refers to the position in which natural spinal curvature and arm, shoulder, hip, knee position is maintained. Incorrect posture gradually decreases the flexibility of the spine, causes pain in neck, shoulder and back. Furthermore, it has detrimental effects on digestion and respiration. After sitting for a long duration, posture becomes inappropriate reluctantly. 

% This project helps a person to maintain the correct posture by monitoring in real-time and alarming the person whenever he sits in a wrong way.\\
%Back pain is recognized as a common health issue. Almost 75-85\% of adults deal with back pain for at least once in their lifetime \cite{tanaka2005epidemiology}. Low back pain and neck pain comes in the $4^{th}$ place of DALY (disability-adjusted life after) worldwide in 2016 \cite{hay2017global}. The World Health Organization (WHO) claims that 9.5 million people in South-East Asia are affected by back and neck pain \cite{hameed2013prevalence}. A study finds that 46\% of school teachers in Dhaka city have reported about being impacted by back pain \cite{akter2018prevalence}. Lack of knowledge of correct posture is assumed to be the primary cause of backache, yet a study finds 60.8\% of Bangladeshi physiotherapists have low back pain \cite{mondal2018prevalence}. In another study, 24.7\% of female Ready-Made Garment (RMG) employees of Bangladesh have reported neck pain \cite{hossain2018prevalence}.
Proper sitting posture refers to the position in which natural spinal curvature and arm, shoulder, hip, and knee position are properly maintained and aligned. Incorrect posture gradually decreases the flexibility of the spine causing pain in neck, shoulder and back. Furthermore, incorrect sitting posture has detrimental effects on digestion and respiration. Back pain is recognized as a common health issue and 75-85\% of adults deal with back pain at least once in their lifetime \cite{tanaka2005epidemiology}. In 2016, low back pain and neck pain combinedly were the $4^{th}$ leading cause of disability-adjusted life year (DALY) globally \cite{hay2017global}. The World Health Organization (WHO) reported that 9.5 million people in South-East Asia were affected by back and neck pain. A descriptive cross-sectional study conducted among school teachers from different school in Dhaka showed 46\% of school respondents have reported lower back pain \cite{akter2018prevalence}. Lack of knowledge of correct posture is assumed to be the primary cause of backache, yet a study that 60.8\% of physiotherapists in Bangladesh have low back pain \cite{mondal2018prevalence}. A cross-sectional study on Ready-Made Garment employees (RMG) of Bangladesh have reported  24.7\% of female worker respondents had lower back pain and 23.\% of female worker respondents had neck pain \cite{hossain2018prevalence}. As many people spend most of the time sitting down in the workplace or home, bad alignment of sitting causes several problems such as bad support on muscles, ligaments, tendons and unhealthy spine \cite{Southern78:online}.
\\
%As many people spend most of the time sitting down in the workplace or home, bad alignment of sitting causes several problems such as bad support on muscles, ligaments, tendons and unhealthy spine \cite{Southern78:online}. Low back pain, harmful impact on several organs and kyphosis are the results of poor posture \cite{cho2017realigning}.
%For the treatment of these abnormalities, it is advised to stay active and bed rest is often discouraged and a doctor or a physician generally prescribes medication (non-steroidal and anti-inflammatory drugs) \cite{koes2006diagnosis}, physical therapies \cite{rajfur2017efficacy}, exercises \cite{rainville2004exercise} \cite{abdollahzade2017effects}, surgery  \cite{cho2014surgical}, non-medical inventions \cite{meade1990low}, spinal traction therapy \cite{oh2018impact} etc.
There are several treatment methods for these abnormalities and depending on the severity, doctor advices to stay active and/or bed rest, prescribe medication \cite{koes2006diagnosis}, physical therapies \cite{rajfur2017efficacy}, exercises \cite{abdollahzade2017effects}, surgery \cite{cho2014surgical}, spinal traction therapy \cite{oh2018impact} etc. On the other hand, as a proactive practice, good sitting posture improves the body's circulatory and digestive system, increases self-confidence, decreases the risk of abnormal wearing of the joint surfaces, improves core and scapular strength etc \cite{12Benefi52:online}. Therefore, it is an excellent decision to make a habit of good sitting posture by monitoring as it brings many benefits for the human body without any unnecessary side-effects. There has been a lot of effort to prevent back pain using eTextile sensor \cite{xu2011ecushion} and flex sensor to monitor sitting posture based ergonomic design. Moreover, there are some approaches based on machine learning and accelerometers \cite{roh2018sitting} \cite{wong2008detecting}. Most of these are very useful but complex, difficult to operate and very expensive. 
\\ 
%On the other hand, good sitting posture improves the circulatory and digestive system of the body, increases self-confidence, decreases the risk of abnormal wearing of the joint surfaces, improves core and scapular strength etc \cite{jonaitis_2053}. So, it is an excellent decision to make a habit of good sitting posture by monitoring as it brings lots of benefits for the human body.
%There has been a lot of researches in this field and most of the time sensors like eTextile sensor \cite{xu2011ecushion}, flex sensor are heavily used in monitoring sitting posture.
% \cite{HumanPos23:online}.
%Also some approaches based on machine learning, accelerometers, radiographical \cite{roh2018sitting} \cite{wong2008detecting} \cite{Wong_Wong_2008} etc, have been used. Some of these are very useful but complex and not user friendly. Besides there is also commercial smart office chair designed to monitor posture which is available on the market, but its too costly \cite{Paginani89:online}.
The objective of this project is to develop a relatively inexpensive, comfortable and user-friendly system to monitor sitting posture in real-time. An elementary portable device connected to a smartphone via bluetooth using a flex sensor is represented. The user has to set the range in which he feels comfortable and receives a vibration in the motor when the angle crosses the safe range of good posture.    
%Therefore, main purposes of this project are to make a low cost, comfortable and user-friendly system. Here, an elementary, mobile device using a flex sensor and Arduino connected with a smartphone via Bluetooth is represented, so any user can monitor sitting posture via a notification on the smartphone. Characterizing the best and safe range of values from the flex sensor attached to the back-support belt that changes its value with changing of the angle of sitting position regarding various professions and types of people like older adults, pregnant women etc. It will help the user to get good sitting posture habit. The user has to set the range in which it feels comfortable. When sitting angle gives sensor value out of that specific range, the smartphone app will give a notification and a vibration in motor.   
%The rest of this paper is categorized in Section II represents the methods that contain instrumentation, overall design and study analysis, in Section III, experimental evaluation is described, the discussion of the result, cost analysis, limitations and future prospects are described in Section IV and at last in Section V summary is discussed briefly.
\section{Design Methods}
% \begin{enumerate}
% \item Provides instructions to sit correctly.
% \item Alarms the user whenever posture is incorrect.
% \item Measures the duration of being seated.
% \item Serves as a treatment to the back pain.
% \end{enumerate}
\subsection{Overall Design}
\begin{figure}[h!]

  \centering
  \includegraphics[width=\linewidth,height=5.4cm]{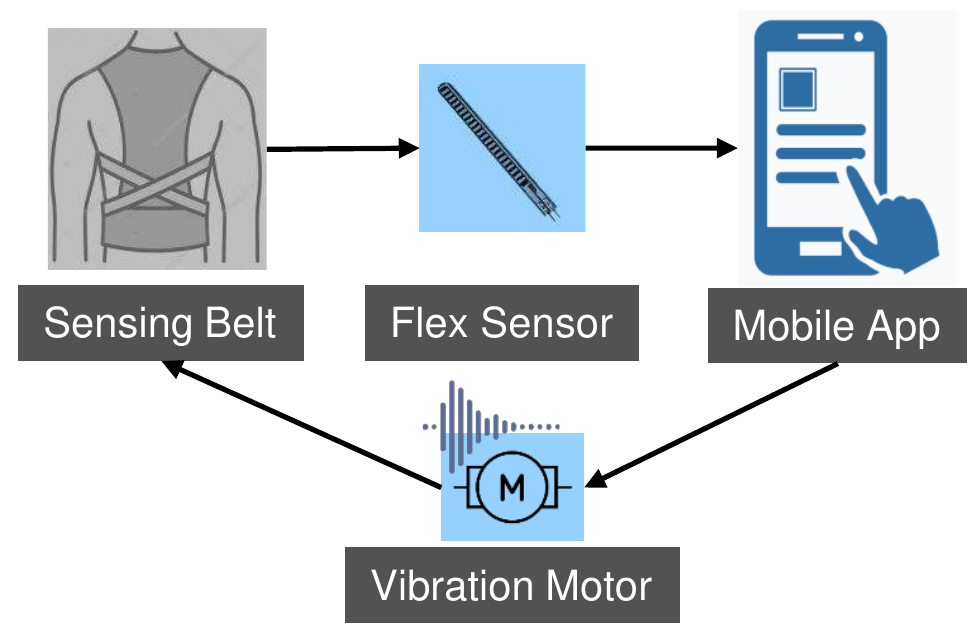}\\
  \caption{Block diagram of the system.}
  \label{fig:block}
\end{figure}

%The proposed design contains two parts; a circuitry part and a mobile application(APP).
%The circuitry part mainly analyzes sitting posture and sends signal to the app. The app monitors the duration of sitting and gives feedback to the user. In Fig. \ref{fig:block} the block diagram represents the system. 
The proposed design contains two parts: a circuitry part and a mobile application (app) part (Fig. \ref{fig:block}).
The circuitry part mainly determines the sitting posture pattern and sends signal to the app. The app monitors the duration of sitting and gives feedback to the user it the bending reaches the safety threshold limits. 

\subsection{Instrumentation}

\subsubsection{Materials}
 %Detachable flex sensor, vibration motor, Bluetooth device are used and an Arduino  nano is used to operate them. All of these electrical devices are assembled on a printed circuit board. A posture supporting belt is used that holds the whole circuitry part mentioned above. For monitoring and feedback, an android application is used. 
Detachable flex sensor (SEN-08606), Vibration Motor (Grove), Bluetooth (HC-05) device are used along with an Arduino-nano (Atmel ATmega 328) to operate and/or control. All of these electrical devices are assembled on a printed circuit board (PCB). A back supporting belt (Royal Posture) is used that holds the whole circuitry part. A 12V rechargeable LIPO battery with average lifespan of 150-250 cycles is used as a power supply. The overall weight of the device including the battery is 340 gm.
%For monitoring and feedback, an android application is used. 
\subsubsection{Circuit Design}
\begin{figure}[h!]
  \centering
  \includegraphics[width=\linewidth,height=6.6cm]{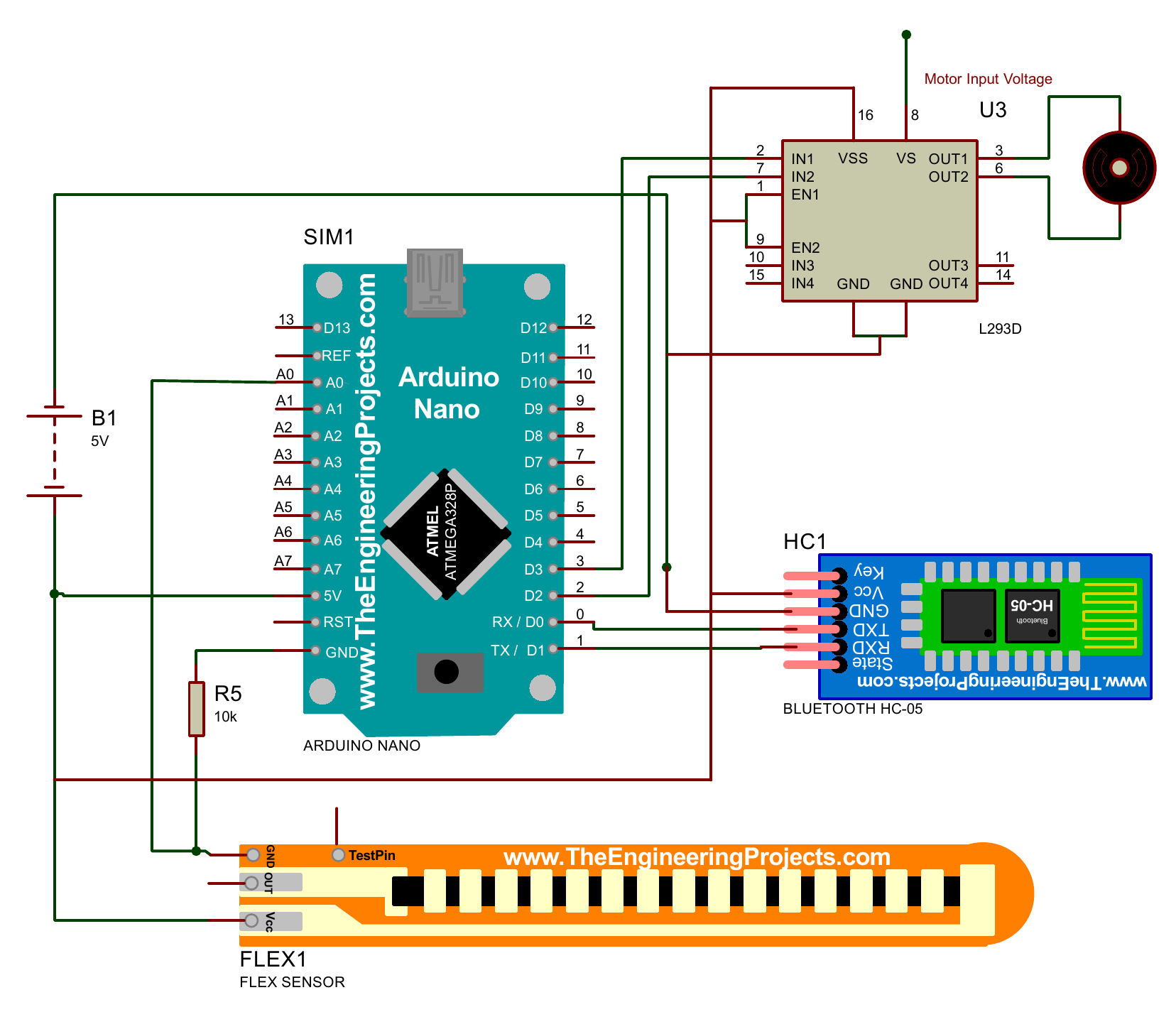}\\
  \caption{Circuit connections in proteus software.}
  \label{fig:pcb1}
\end{figure}

%The circuit components implanted on a 4-inch x 4-inch Printed Circuit Board(PCB) and 4-inch-long flex sensor, are attached to the sensor belt. A 12-volt rechargeable power supply and the vibration motor (to notify the user to correct his posture) are also placed on the belt.
% The circuit contains a voltage regulator system that convert the 12-V to 5-Volt that ensure the proper power supply. 
%Bluetooth transfers the sensor value to the android app and thus the user sets the threshold value and continuously monitors the posture. The system is calibrated with the sensor value and posture angle in such a way that the transferred sensor value symbolizes the postural angle of the user. If this angle crosses the threshold of good sitting posture, the feedback of the vibration motor alerts the user to get back to the accurate postural angle. In Fig. \ref{fig:pcb1} PCB connection and layout design are illustrated.
The circuit components are implanted on a 4-inch x 4-inch PCB and 4-inch-long flex sensor whereas the battery and the vibration motor are attached to the sensor belt (Fig. \ref{fig:pcb1}). The circuit contains a voltage regulator system that converts the 12V to 5V to ensure the proper power supply. Bluetooth transfers the sensor value to the android app to set the threshold value based on the user comfortability and continuously monitors the posture position. The correlation between the sensor value and posture angle had been developed to convert sensor value corresponding to the user's postural angle. The vibration motor feedback system alerts the user to get back to the initial set  postural position once the sitting posture angle crosses the threshold of good sitting posture.

\subsubsection{Mobile APP Design}
\begin{figure*}[h!]
  \centering
  \includegraphics[width=\linewidth,height=6.8cm]{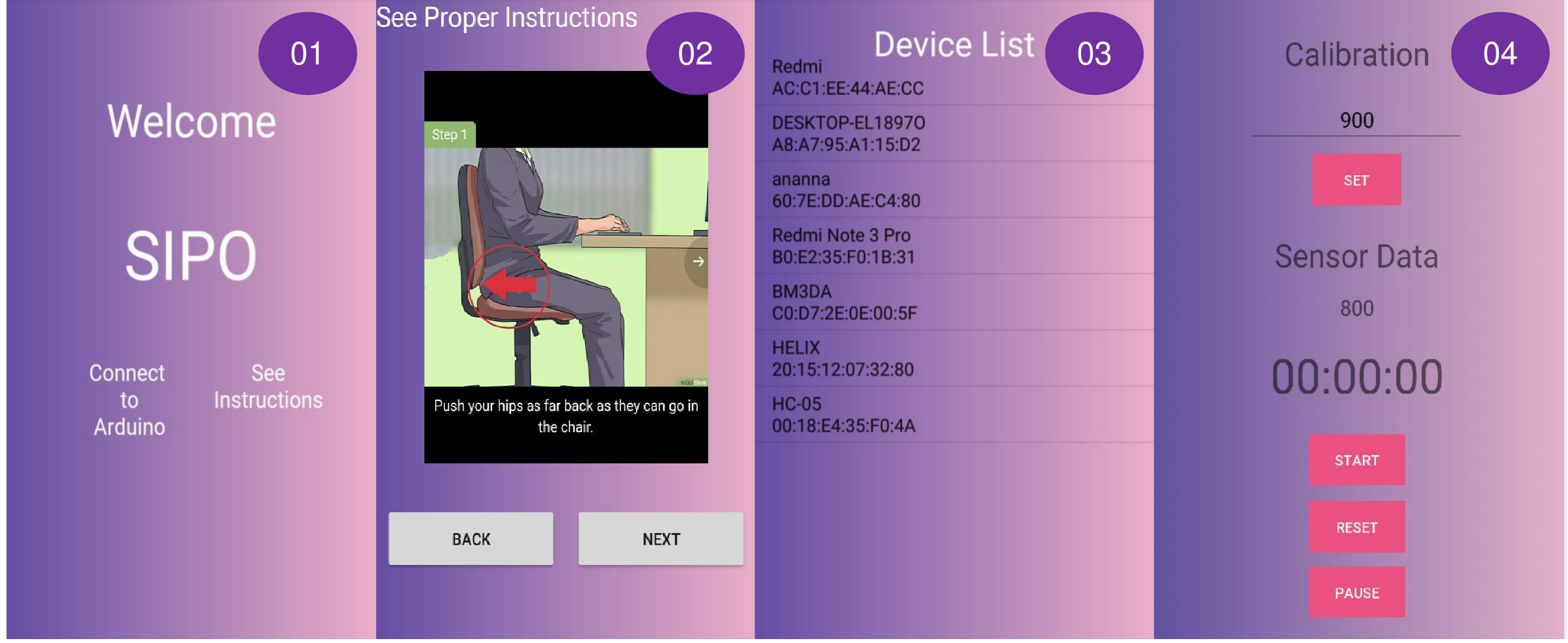}\\
  \caption{Pages of android application, where, (1) Home page, (2) Instruction page, (3) Connection page, and (4) Monitoring and timer tab.}
  \label{fig:app}
\end{figure*}

%The first page of the mobile app named SIPO app has two options i.e., “Connect to Arduino” and “Instructions”. Instruction window illustrates tutorial about the proper sitting process for the user. The user has to connect the app with the Arduino in PCB through connect option. As the process is based on Bluetooth system, device name will appear on the screen. After connecting the device the user has to set a threshold value from the sensor reading on the app screen according to the comfortable position. Then the continuous monitoring of posture is possible. There is a timer feature for counting the total duration of sitting. In Fig. \ref{fig:app} the layout of the android application is provided.
The home page of the mobile app named SIPO app has two options i.e., “Connect to Arduino” and “Instructions”. Instruction window illustrates a tutorial about the proper sitting process for the user (Fig. \ref{fig:app}). The user has to connect the app with the Arduino in PCB through the connect option. After connecting the device, the user has to set a threshold value from the sensor reading on the app screen according to his comfortable position to monitor posture at a fixed position continuously. The timer feature is for counting the total duration of sitting.

\subsection{Study Design}
%An experimental study is done on participants to calibrate the sensor value with posture angle and to get user feedback of the system ensuring participant’s safety and consents. Patient information and spine measurements are taken for further study; a questionnaire proforma is filled for the future development of the system.
An experimental study is done on healthy volunteers having a range of age, weight, height and gender to find the best position for the sensor placement, to calibrate the sensor value with posture angle and to get user feedback of the system to ensure the participant’s safety and effectiveness.

%\subsubsection{Duration of Study}
%For the design and experimental study on participants, at least six months is needed. This time involves the preliminary design, prototype design, final product design and experimental research.

\subsubsection{Settings}
%At first, the setup of the experimental study needs proper space for the comfortable sitting of the participants. Then, a marker is used to observe the movement of the participants and a camera is used to capture for the angle measurement. The mobile app (SIPO) needs to be installed and the circuitry part is used to observe the corresponding sensor value of the posture angle. ImageJ software/ angle meter app is used to calculate the angle from the images. In addition to that, there needs to be some space in the front and back of the participants so that the user can move forward and backward while sitting. Two observers observe the sensor value and one notes it down and the other one observes the angle values using the image. The angle measurement procedure is shown in Fig. \ref{fig:angle}.

\begin{figure}[h!]
  \centering
  \includegraphics[width=\linewidth]{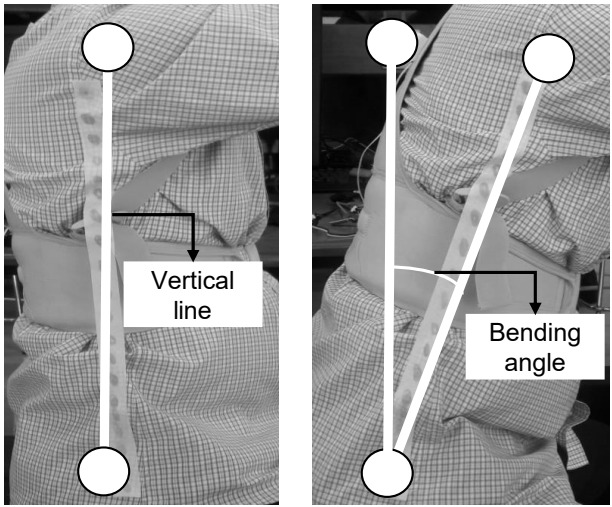}
  \caption{Posture bending angle measurement.}
  \label{fig:angle}
\end{figure}

A marker is used to observe and track  the volunteers bending status when he is advised to bend forward and backward freely from a normal sitting position and a camera is used to capture the position for the angle measurement with respect to the initial position (Fig. \ref{fig:angle}). The mobile app (SIPO) needs to be installed and the circuitry part is used to observe the corresponding sensor value of the posture angle. ImageJ software/ angle meter app is used to calculate the angle from the images. It is important to make sure that there are enough spaces so that the user can easily move forward and backward while sitting.

\subsubsection{Data Collection from Participants}
%To collect data from the participants, four positions are selected to specify the best location for the placement of the sensor. They are- (a) upper thoracic (6 cm below neck), (b) lower thoracic (12 cm below neck), (c) upper lumbar (20 cm below neck), and (d) lower lumbar (26 cm below neck). To measure the relationship between the sensor value and posture angle, participants change their posture angle by moving forward and backward six times while sitting. While movement, the sensor value, and the corresponding angle are measured. A graphical representation of sensor value and angle is used for threshold range. The placement of the sensor is an important factor as it varies with the height of the person. This experimental study helps to generalize the best position of the spine for the placement of the sensor.
Four positions are selected to specify the best location for the sensor placement based on the average human anatomical position to best  track the bending movement. They are- (a) upper thoracic (6 cm below neck), (b) lower thoracic (12 cm below neck), (c) upper lumbar (20 cm below neck), and (d) lower lumbar (26 cm below neck). In order to develop the relationship between the sensor value and posture angle, participants are advised to change their posture angle by moving forward and backward six times (three times each side covering a wide range of posture angle) from sitting at a normal position. The position of the sensor is an important factor as it varies with the height of the person. There are in total 9 participants, age ranging from 21 to 50 ($\pm {7.91}$),  height and weight ranging from 5’ to 5’11’’ ($\pm {0.29}$) and 45 kg to 75 kg respectively.
\subsection{User Feedback}
User feedback form has been filled by the user to evaluate the different aspects of the device from end-user point of view (comfortability, preference, effectiveness, satisfaction). 
%Age and height are major factors in this experimental analysis. There are variations in the relation of sensor value and angle with different age and height of the people. There are in total 9 participants, age ranging from 21 to 50 ($\pm {7.91}$),  height and weight ranging from 5’ to 5’11’’ ($\pm {0.29}$) and 45 kg to 75 kg respectively. \\
%A generalization study is done on range of people considering a specific age and height.
%User feedback form is used where the device is given marks in different aspects (comfortability, preference, effectiveness, satisfaction) by different users. In this case, while acquiring the data from the user, the device is used by the user and an instant feedback is provided comparing existing technologies and price.

\section{Results Analysis}
%The placement of the sensor is a matter of concern to design the device. Analysis on the position variability with the sensor value and threshold calculation is necessary to remove the constraints related to the placement of the sensor. Position variability analysis and threshold calculation are discussed in the following subsections.

\subsection{Position Variability Analysis}
\begin{figure}[h!]
   \centering
  \includegraphics[width=\linewidth]{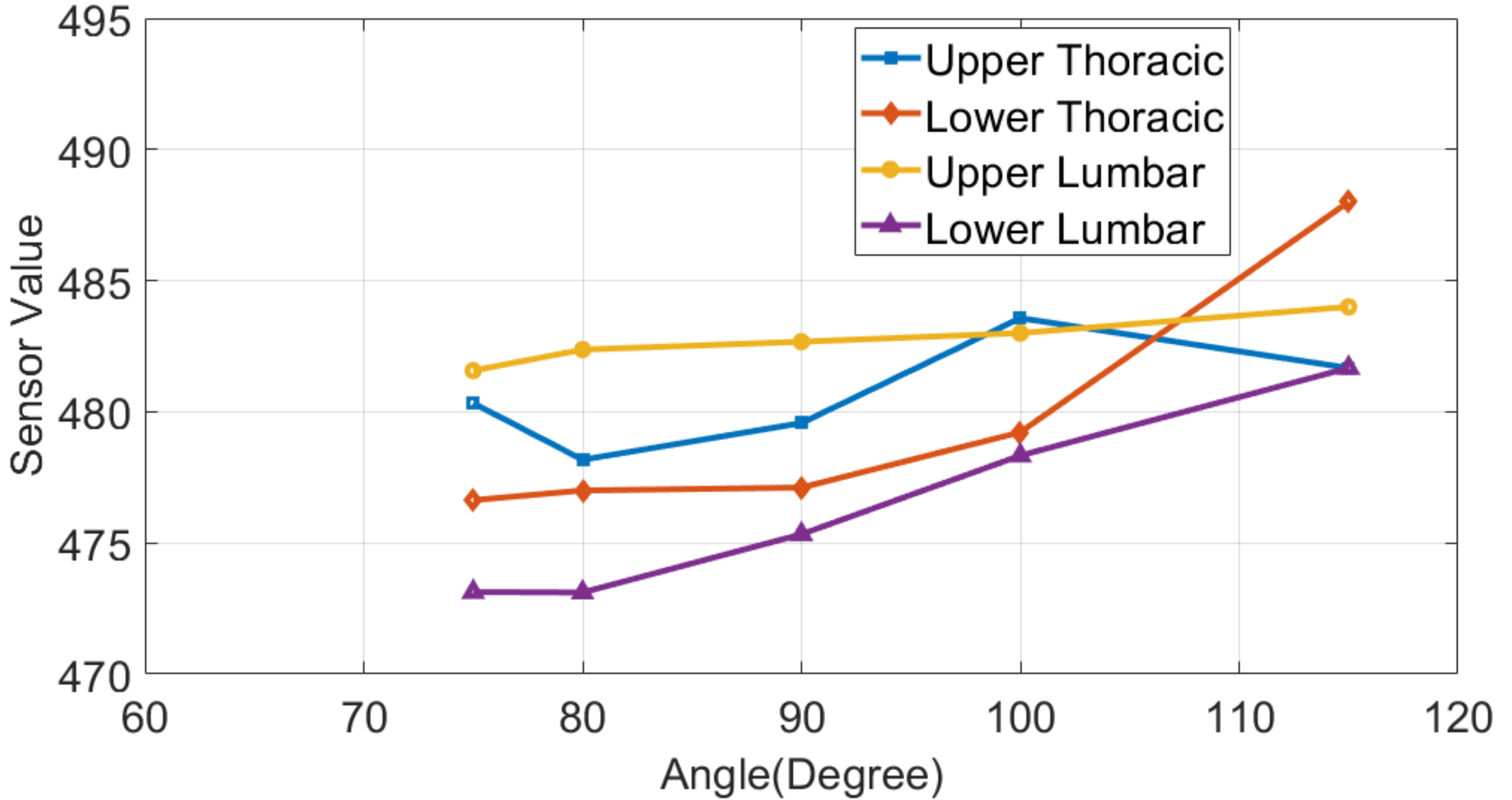}\\
   \caption{Determination of the sensor position for the best sensitivity.}
   \label{fig:sensorplot}
 \end{figure}

%For the analysis of the position variability, the collected data for four different positions are studied. Five different angles are selected ( 75$^{\circ}$, 80$^{\circ}$, 90$^{\circ}$, 100$^{\circ}$, 115$^{\circ}$). For a fixed position and angle value, the sensor value of 9 subjects are averaged and plotted in a graph. In Fig. \ref{fig:sensorplot}, the average curves for four different positions are shown. In lower thoracic position the range of sensor values for different angles is higher, which proves better sensitivity of sensor at this position. For this reason, lower thoracic position is selected for the placement of the sensor, removing user variability. 
For the analysis of the sensor position variability, the collected data of four different positions are studied at five different selected angles ( 75$^{\circ}$, 80$^{\circ}$, 90$^{\circ}$, 100$^{\circ}$, 115$^{\circ}$). For a fixed position and angle value, the sensor values of 9 subjects are averaged (Fig. \ref{fig:sensorplot}). In the lower thoracic position the range of sensor values for different angles is higher, which shows better sensitivity of sensor and wide range of sensor values at this position. For this reason, the lower thoracic position is selected for the placement of the sensor. 

\subsection{Threshold Range Calculation}
\begin{figure}[h!]
   \centering
  \includegraphics[width=\linewidth]{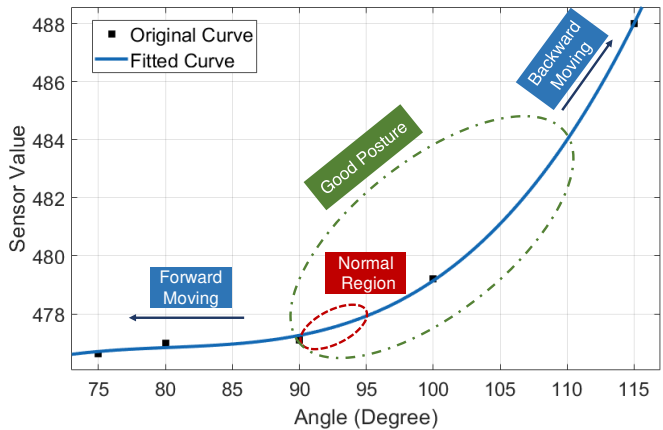}\\
   \caption{Calibration between the sensor value and bending angle.}
   \label{fig:11}
 \end{figure}

%To find the equation for the calculation of threshold range, the curve in Fig. \ref{fig:11} for lower thoracic position is fitted with a $3^{rd}$ degree of polynomial equation which is shown in Eqn. \ref{eqn:1}. An increase in the sensor value with the increase of angle is observed because of the bending. Using the derived equation, the deviation of sensor value for the good posture angles is measured and a threshold range for good posture is set to get the feedback from the device. 
Bending angle range of 90$^{\circ}$ - 110$^{\circ}$ has been reported as the safe zone for good posture whereas 90$^{\circ}$ - 95$^{\circ}$  is the range  of normal sitting posture (Fig. \ref{fig:11}) \cite{ProperSi99:online}. Therefore, in order to develop the corresponding sensor value for good posture position, a model has been developed for lower thoracic position and fitted with a $3^{rd}$ degree of a polynomial equation (Eqn. \ref{eqn:1}). So, sensor value beyond the thresholds of safety regions generates signal to the vibration motor and alert the user to change the sitting position.   

% \cite{grandjean1977ergonomics}
\begin{equation}
SV=0.0003(A)^3-0.0605(A)^2+4.8789(A)+345.23
\label{eqn:1}
\end{equation}
where SV = sensor value and A = angles in degrees.
%LT_2n.pdf

\section{Discussion}
The survey on the users' feedback gets 7.6/10 in terms of comfortability, 9.2/10  in terms of preference, 9.6/10 effectiveness and 8.0/10 in terms of overall satisfaction  based on the average of the given marks by the users in user feedback form. The feedback also indicates that the user will be able to use this device for 5-6 hours a day. \\
Continuous monitoring of a posture position through the developed app makes the device more used-friendly, comfortable, portable along with low-cost compared to the other devices available in the market. The available devices in the market cost about 150 USD - 600 USD, whereas the cost of SIPO is only about 42 USD. The manufacturing cost can be reduced by mass production thus making it an affordable solution for those who prefer to continuously monitor their sitting posture and achieve a healthy spine.\\ 
Sensitivity and accuracy of the system can be upgraded in future by controlled movement of bending during the experiment using more sophisticated camera and automatic calculation of the bending angle. More data from a large population having wide variation of height, weight, and age can be used to calibrate the sensor value with the bending angle. Moreover, the apps can be updated by adopting artificial intelligence and determine the initial set position and threshold values based on the user inputs (weight, age, gender, and height).

\section{Conclusion}
%The primary goal of this project is to provide a pleasant solution to maintain a perfect sitting posture which will be more user-friendly, inexpensive, non-invasive and beneficial to health for users. Using modern technology and app system, this project provides a cost-effective way for real-time posture monitoring of a seated person. Due to lightweight and portability, this wearable product is one of the best solutions for the desk-workers, students and travelers who spend a significant amount of time sitting. 
The primary goal of this project is to provide an effective solution to maintain a healthy sitting posture that is user-friendly, wearable,  safe, inexpensive and non-invasive. The developed functional prototype demonstrates a cost-effective way for real-time monitoring of a posture position. Due to light-weight and portability, this wearable product is one of the best solutions for the desk-workers, students and travelers who spend a significant amount of time in sitting position.

\section{Acknowledgment}
This work has been conducted with the help of Department of Biomedical Engineering, Bangladesh University of Engineering and Technology. The authors would like to thank the volunteers for their cooperation. Special thanks to Navid Ibtehaj Nizam, Assistant Professor, Department of Biomedical Engineering, Bangladesh University of Engineering and Technology.

\bibliographystyle{./bibliography/IEEEtran}
\bibliography{bibtex.bib}

\end{document}